\documentclass{appolb}
\usepackage{graphicx}
\usepackage{amsmath}

\def\gs{g_{\mathrm{s}}}

\def\mD{m_{\mathrm{D}}}

\def\bbp{\bb_\perp}
\def\bbp{b_\perp}
\def\bbps{b_\perp^2}

\def\qp{q_\perp}
\def\bqp{\boldsymbol q_\perp}

\def\Cbp{C(\bbp)}
\def\Cqp{C(\qp)}

\def\d{\mathrm{d}}
\def\e{\mathrm{e}}

\begin{document}
\title{ Non-perturbative determination of the collisional broadening kernel and medium-induced radiation in QCD plasmas%
\thanks{Presented at 29TH INTERNATIONAL CONFERENCE ON ULTRA – RELATIVISTIC NUCLEUS - NUCLEUS COLLISIONS}%
}

\author{S\"oren Schlichting
\address{Fakult\"at f\"ur Physik, Universit\"at Bielefeld\\ 
D-33615 Bielefeld, Germany.}
\\[3mm]
{\underline{Ismail Soudi}\footnote{Speaker.}
\address{Department of Physics and Astronomy, Wayne State University\\
Detroit, MI 48201.}
}
}

\maketitle
\begin{abstract}
Collisional broadening in QCD plasmas leads to the emission of medium induced radiation, which governs the energy loss of highly energetic particles or jets. While recent studies have obtained non-perturbative contributions to the collisional broadening kernel $C(b_\bot)$ using lattice simulation of the dimensionally reduced long-distance effective theory of QCD, Electrostatic QCD (EQCD), so far all phenomenological calculations of jet quenching rely on perturbative determinations of the collisional broadening kernel. By matching the short-distance behavior of the lattice extracted EQCD broadening kernel, we determine the fully matched QCD broadening kernel non-perturbatively. We present results for the collisional broadening kernel in impact-parameter ($C_{\rm QCD}(b_\bot)$)) and momentum space ($C_{\rm QCD}(q_\bot)$) and employ them to determine the rates of medium induced radiation in infinite and finite size QCD plasmas. By contrasting our results with leading and next-to-leading order perturbative determinations as well as various approximations of the splitting rates employed in the literature, we investigate the effect of the non-peturbative determination of CQCD($q_\bot$) on medium-induced radiation rates.

\end{abstract}
  
\section{Introduction}
During ultra-relativistic heavy-ion collisions, high energetic particles originating from the initial hard partonic collisions, must traverse and interact with the Quark-Gluon-Plasma (QGP) before reaching the detector. 
Interactions of highly energetic parton with the medium have been studied extensively, and it was shown that besides the elastic interactions with the medium constituents, multiple soft scatterings between the hard partons and the medium trigger the partons to radiate. 
These soft scatterings lead to an infinite number of diagrams, which can be resummed to obtain an effective $1\to2$ in-medium radiation rates. Several formalisms for in-medium radiation have been developed in the literature \cite{Gyulassy:1999zd,Arnold:2002ja,CaronHuot:2010bp}, which encapsulate the same physics but differ on the level of approximations. However, all these formalisms rely on a description of the elastic scattering with the medium, obtained using the transverse momentum broadening kernel 
\begin{equation}    \label{C_q_perp}
    \Cqp \equiv \frac{(2\pi)^2 \d^3 \Gamma}{\d^2 \qp \, \d L} \, ,
\end{equation}
which describes the rate at which the hard partons exchange transverse momentum $\bqp$ with the medium. 
To obtain the broadening kernel three treatments of the medium are widely used in the literature : \\
    \textit{Many random, static, screened color centers} \cite{Gyulassy:1999zd}: $\Cqp \propto \frac{1}{(\qp^2 + \mD^2)^2}$.\\
    \textit{Dynamical moving charges} at lowest order in perturbation theory \cite{Aurenche:2002pd}: $\Cqp \propto \frac{1}{\qp^2(\qp^2+\mD^2)}$.\\
    \textit{Many individually small scatterings}, leading to transverse momentum diffusion \cite{Arnold:2008iy}: $\Cbp = \hat{q} \, \bbps/4$.\\
Recently, there have been progress in obtaining non-perturbative contributions to the broadening kernel \cite{Moore:2019lgw,Moore:2021jwe}, using a dimensionally reduced long-distance effective theory for QCD, 3D Electrostatic QCD (EQCD) \cite{CaronHuot:2008ni}. After a Fourier transform of the broadening kernel to impact parameter space, lattice simulations of EQCD \cite{Panero:2013pla,DOnofrio:2014mld,Moore:2019lgw} can be performed to obtain the large distance behavior of the broadening kernel $C_\mathrm{EQCD}(\bbp)$, which was matched to the short distance behavior of QCD to obtain a non-perturbative determination of $C_\mathrm{QCD}(\bbp)$ in QCD at all scales \cite{Moore:2021jwe}. 
During these proceedings, we employ the broadening kernel extracted to obtain the in-medium splitting rates of a highly energetic parton traversing a medium with a finite extent and compare with results obtained using perturbative results of the broadening kernels. 
\section{Non-perturbative broadening kernel}
\begin{figure}
    \centering
    \includegraphics[height=0.35\textwidth]{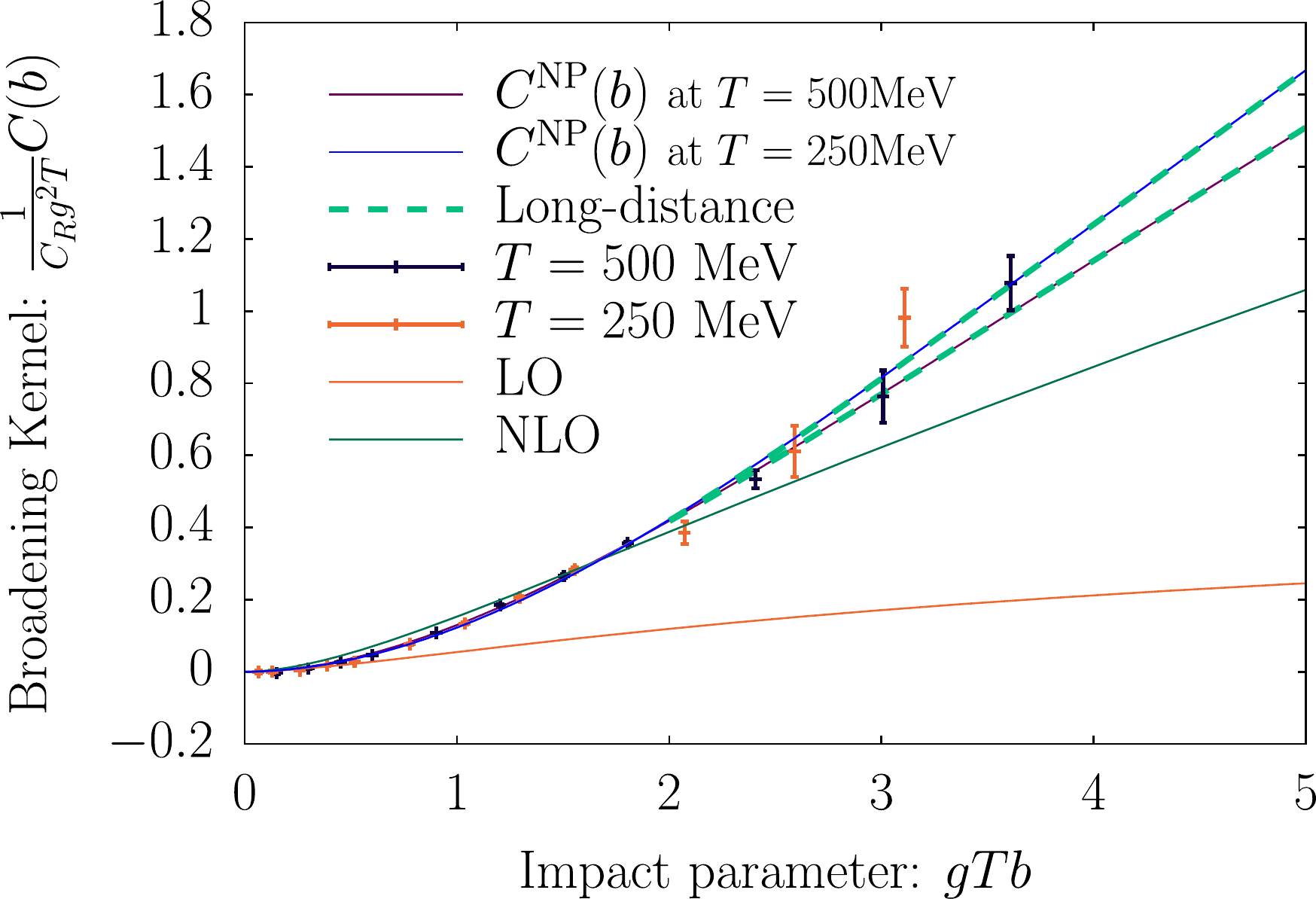}
    \includegraphics[height=0.38\textwidth]{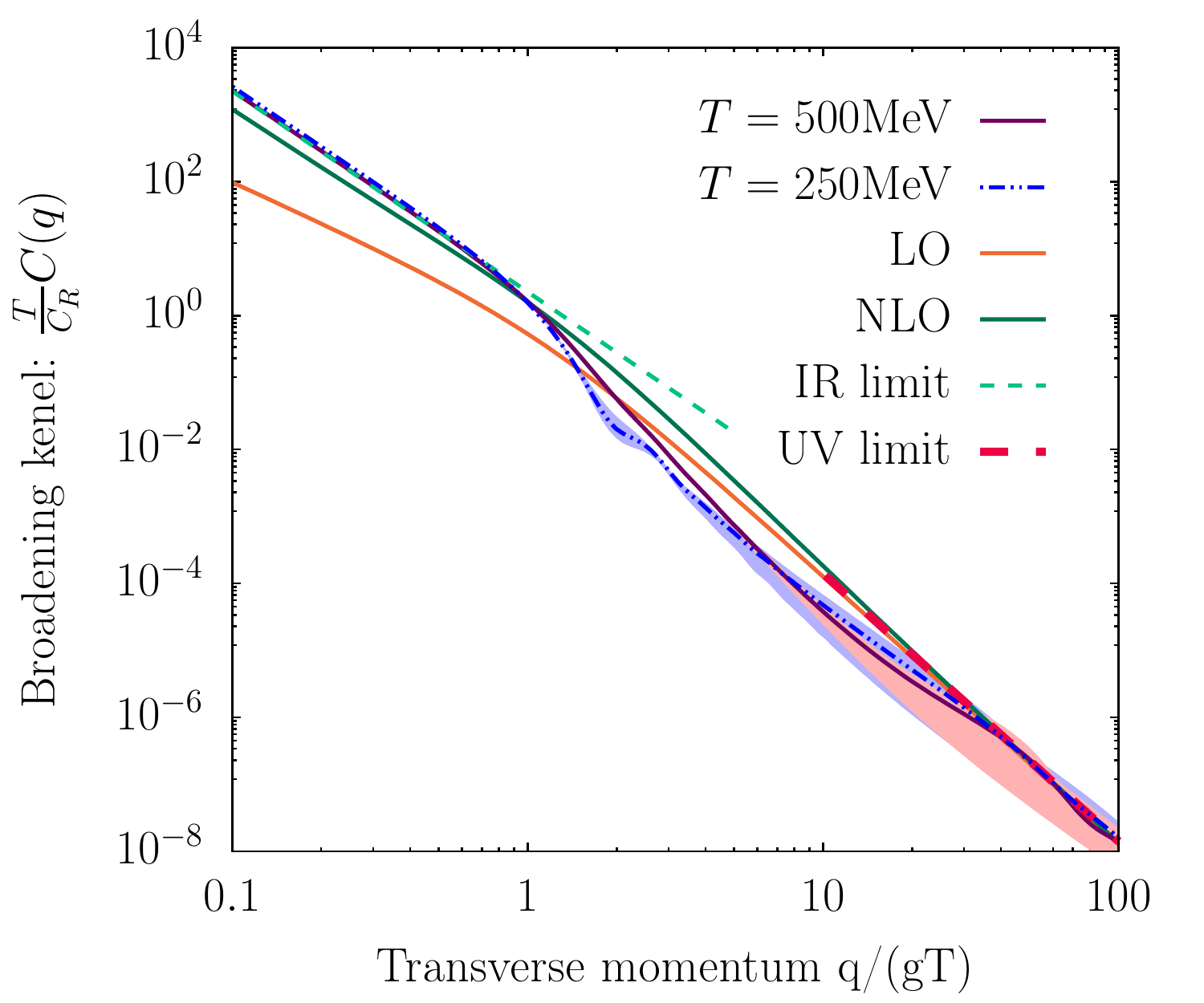}
    \caption{(left) Non-perturbative elastic broadening kernel $C_{\rm QCD}(\bbp)$ in impact parameter space.  Data points for two different temperatures $T=250,500 {\rm MeV}$ are shown alongside the interpolating splines. (right) Elastic broadening kernel $C_{\rm QCD}(\bqp)$ in momentum space for $T=250,500$MeV. We also compare the kernel to leading-order (LO) and next-to-leading order (NLO) determinations at $T=500$MeV. }
    \label{fig:InverseKernel}
\end{figure}
The non-perturbative (NP) broadening kernel $(C_\mathrm{QCD}(\bbp))$ in impact parameter space $\bbp$ was obtained in \cite{Moore:2021jwe}, and shown in the left panel of Fig.~\ref{fig:InverseKernel}. 
However, in order to compute in-medium splitting rates in a QCD medium of finite size, it is more preferable to work in momentum space. Therefore, we perform a Fourier transform to obtain the non-perturbative broadening kernel $(C_\mathrm{QCD}(\qp))$ in impact parameter space $\qp$ which is shown for $T=250,500$MeV in the right panel of Fig.~\ref{fig:InverseKernel} (c.f. \cite{Schlichting:2021idr} for details of the procedure).
We find that both data sets at $T=250,500$MeV display very similar behavior when expressing $T C_\mathrm{QCD} (\qp)/C_R $ as a function of $\qp/\gs T$ in momentum space, or $C(\bbp)/\gs^2TC_R$ as a function of $\gs T \bbp$ in impact parameter space.
When comparing with leading order (LO) and next-to-leading order (NLO) pertubative broadening kernels, we find that while they all follow the same ultraviolet (UV) behavior $ \sim 1/\qp^4$ and respectively $ \sim \bbp^2\log(\bbp)$ behavior, the infrared (IR) behavior of the NP and NLO kernels is markedly different from the LO. The NP and NLO kernels in UV follow a $\sim1/\qp^3$ and respectively $\sim\bbp$. However, the slopes of the kernels differ by a prefactor due to the difference in the string tension $\sigma_{\rm EQCD}$.

\section{In-medium splitting rates}
\begin{figure}
    \centering
    \includegraphics[width=0.48\textwidth]{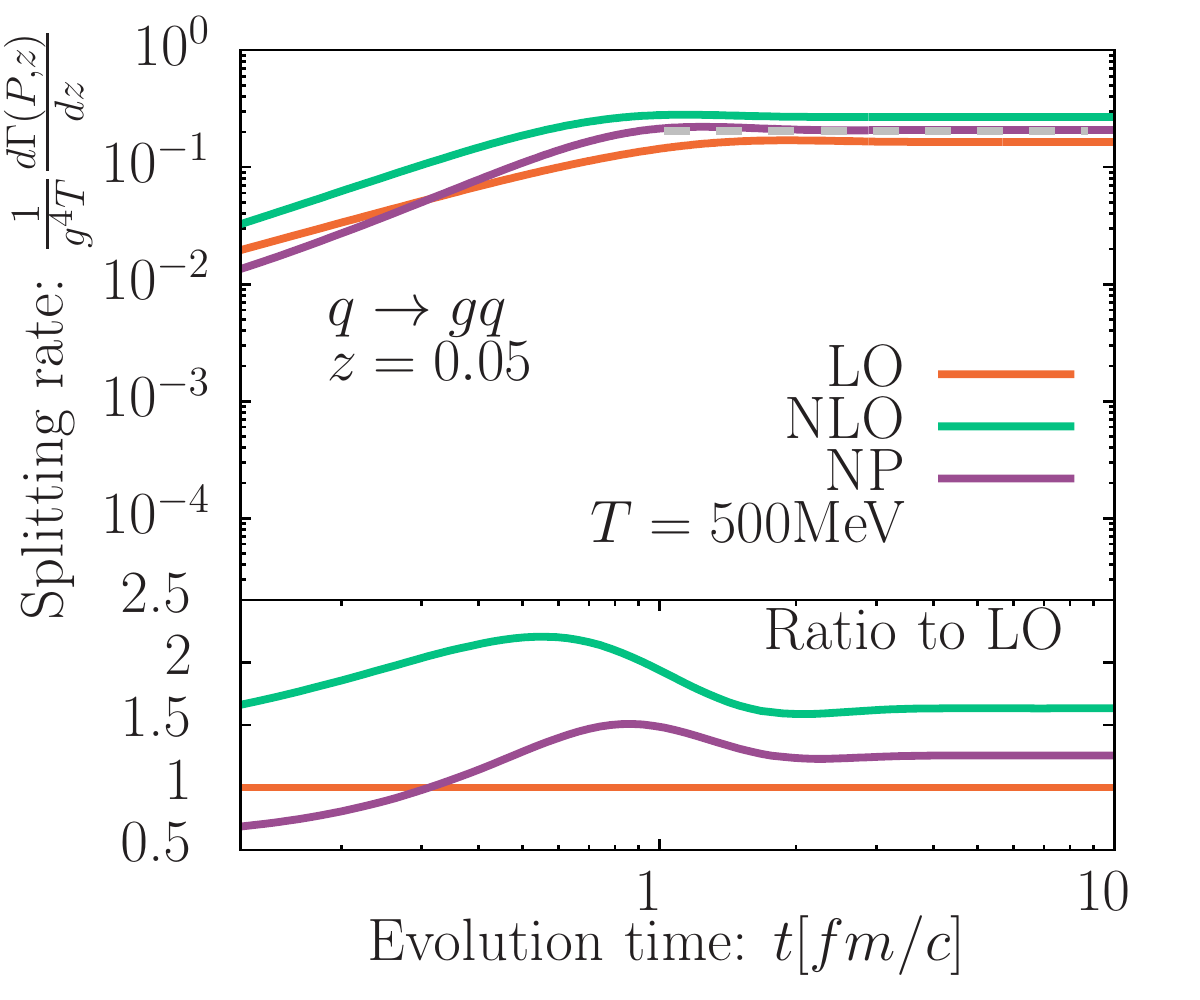}
    \includegraphics[width=0.48\textwidth]{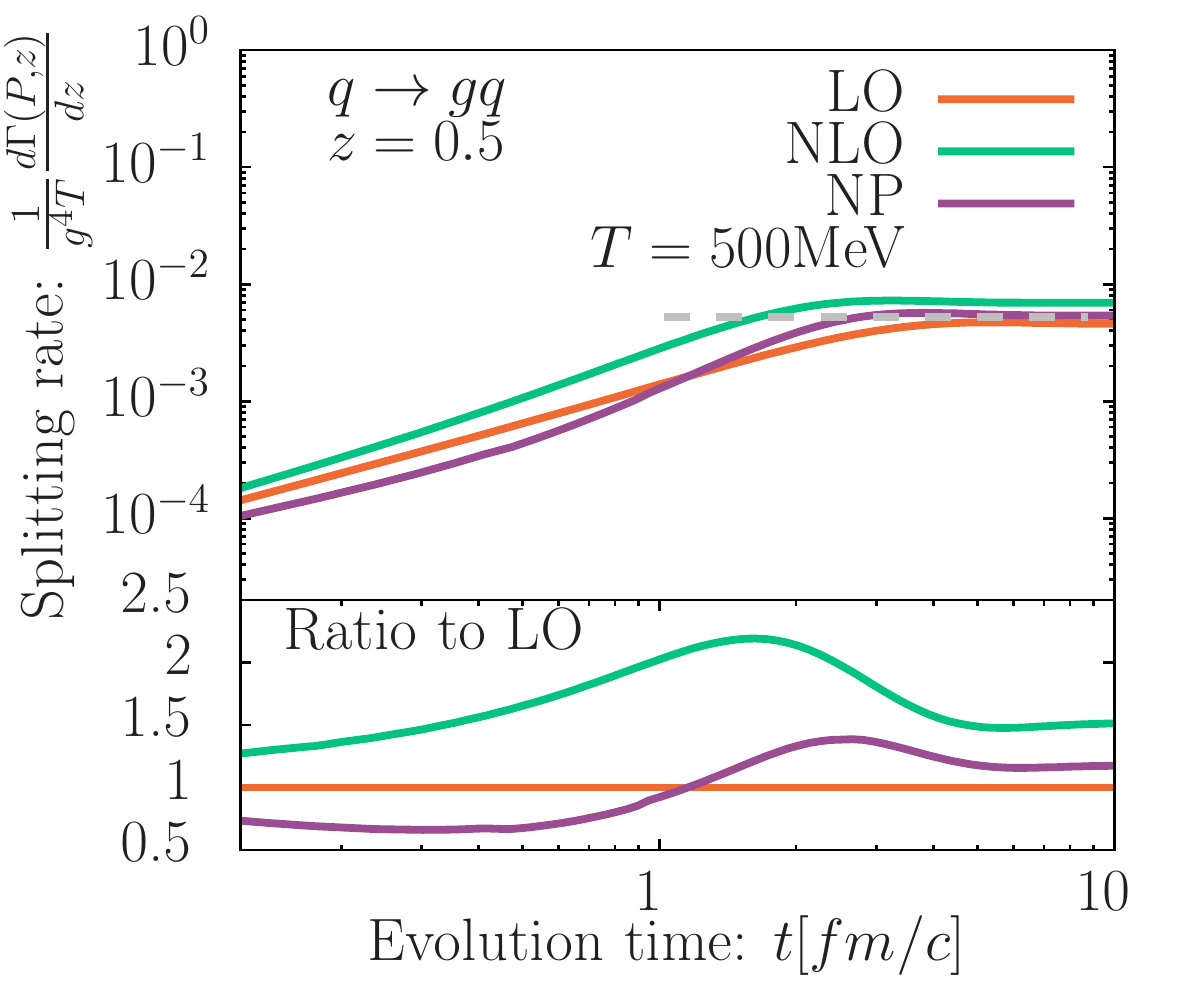}
    \includegraphics[width=0.48\textwidth]{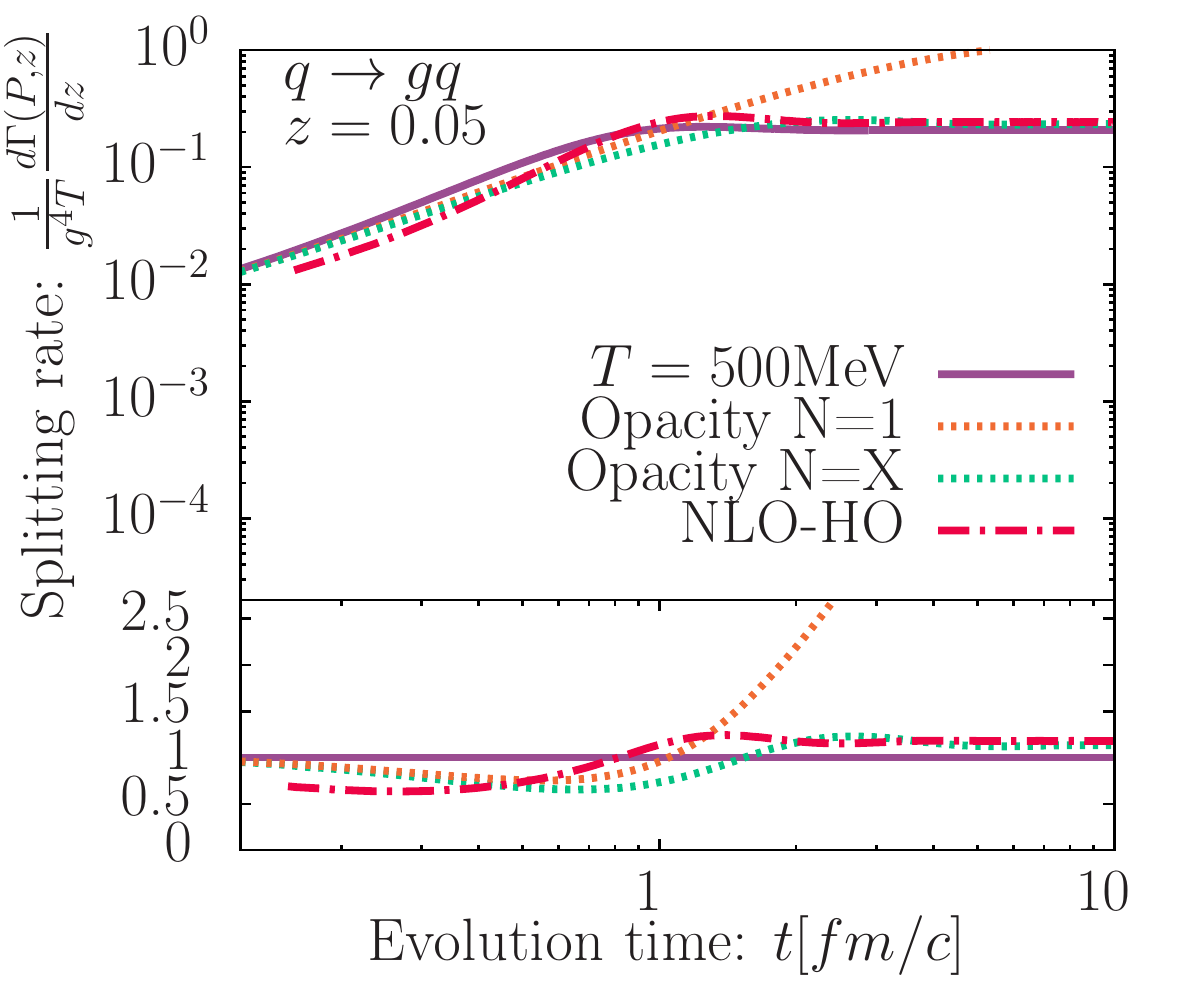}
    \includegraphics[width=0.48\textwidth]{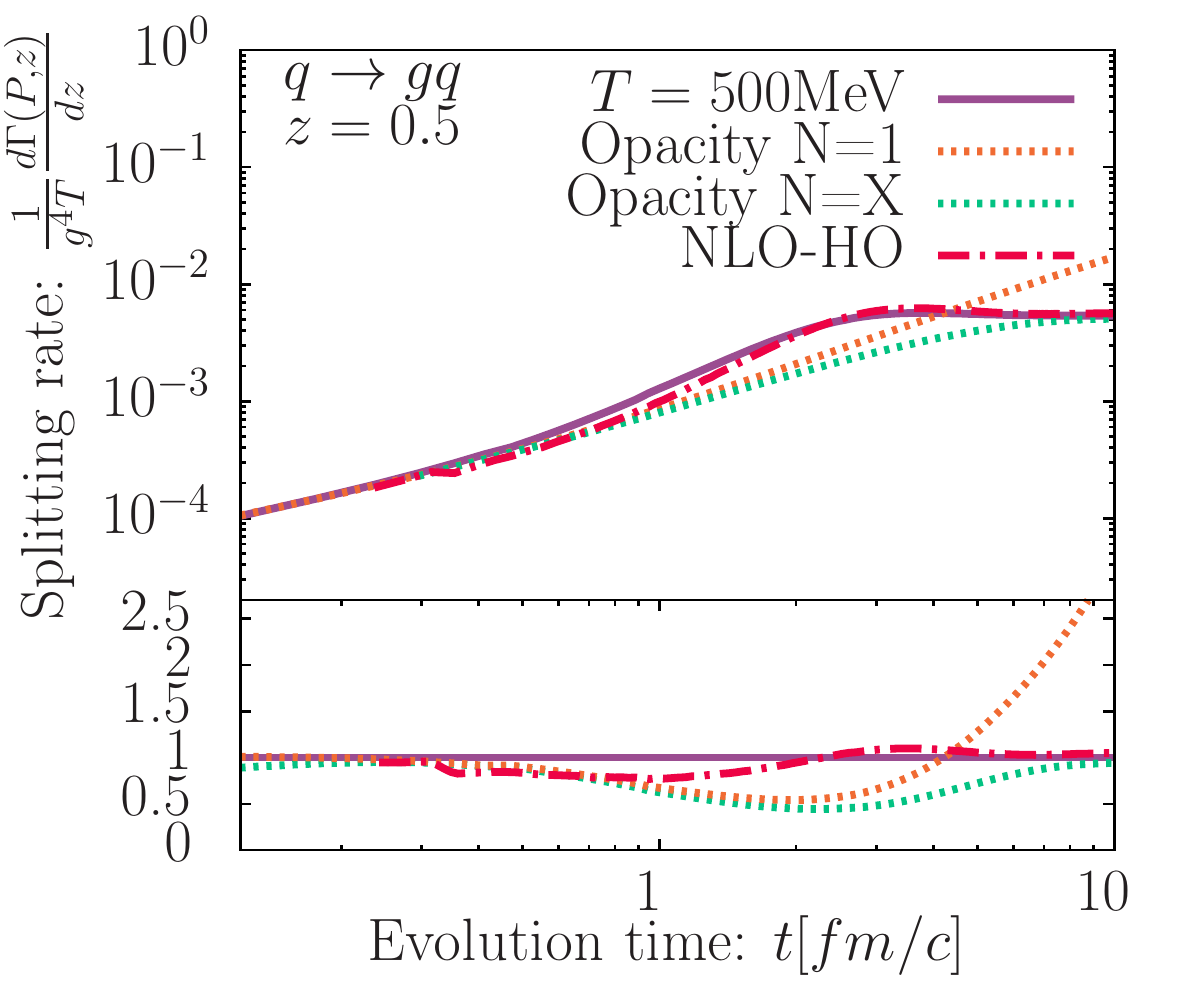}
    \caption{Splitting rate for the medium-induced emission of a gluon from a parent quark with energy $P=300T$ in an equilibrium plasma with temperature $T=500$MeV as a function of the evolution time $t$ for different gluon momentum fraction $z=0.05,0.5$. (top) Different curves in each panel show the results for the different, leading order (LO), next-to-leading order (NLO) and non-perturbative (NP) momentum broadening kernels in Fig.~\ref{fig:InverseKernel}. 
    (bottom) We compare different approximations of the in-medium splitting rate, namely the opacity expansion at $N=1$ \cite{Gyulassy:1999zd}, the resummed opacity rate of \cite{Andres:2020kfg} ($N=X$) and the NLO expansion around the Harmonic Oscillator \cite{Mehtar-Tani:2019tvy} (NLO-HO) to the full result ($T=500{\rm MeV}$). Note that all results are obtained with the non-perturbative collision kernel. The lower panel of each plot displays the ratio to the full rate.}
    \label{fig:FiniteMediumVSKernels}
\end{figure}

Employing the different broadening kernels, and following \cite{CaronHuot:2010bp}, we obtain the rate of medium-induced radiation $\frac{d\Gamma(P,z;t)}{dz}$, the rate of quark with momentum $P$ radiating a gluon with momentum $zP$ after a time $t$ spent in the medium. The results for momenta fraction $z=0.05$ and $z=0.5$ are shown in Fig.~\ref{fig:FiniteMediumVSKernels} as a function of time $t$ where we used the LO, NLO and NP broadening kernels to obtain the orange, green and purple curves, and we show the ratio to the LO curve in the lower inset of each panel. 
Initially we observe a linear behavior of the rate which saturates rather quickly at later times and recovers the infinite medium rates which were obtained earlier~\cite{Moore:2021jwe}.
We find that the non-perturbative result starts lower than the LO rates before it settles above the LO and below the NLO, which can be explained by the fact that the early times radiation are driven mainly by single hard scatterings with large momentum exchange $(\qp\gg m_D)$ where the NP kernel is below the LO. At late times, multiple soft scatterings with small momentum exchange $(\qp\sim m_D)$ are more important, and the NP kernel in this IR behaves similarly to the NLO kernel. Overall, the results obtained using the NP kernel do not depart from a band of $\pm50\%$ around LO, while the NLO results can become over $2\times$ larger than the LO result. 

Additionally, we computed the in-medium radiation rates using several approximations developed in the literature:
    \textit{First order opacity expansion}, which considers a perturbative expansion in the number of re-scatterings with the medium. Here we take only a single scattering ($N=1$) \cite{Gyulassy:1999zd}. \\
    \textit{Resummed opacity expansion} ($N=X$) which re-sums the opacity expansion by employing a cut-off to the momentum exchange of the scattering, the collision integral of subsequent scatterings can then be exponentiated (c.f. \cite{Andres:2020kfg}).\\
    \textit{Harmonic Oscillator expansion} (NLO-HO) where using an opacity expansion, a single hard scattering is considered as a perturbation on top of fully resummed multiple soft-scatterings computed in the diffusion approximation \cite{Mehtar-Tani:2019tvy}.\\
Extensive details of calculation of each approximation are given in \cite{Schlichting:2021idr}. We employ these approximations to obtain the in-medium splitting rate for the same NP broadening kernel $C_{\rm QCD}(\qp)$ at $T=500$MeV, and show a comparison in Fig.~\ref{fig:FiniteMediumVSKernels}. While the naive first order opacity expansion reproduces the linear behavior of the rate at early times, it breaks down later on when the time is large enough to allow for subsequent scatterings. Conversely, the resummed $(N=X)$ opacity expansion is able to reproduce the rate even at late times. For the NLO expansion around the HO, we observe that it can be quite accurate throughout the evolution especially for quasi-democratic splittings ($z\sim 1/2$). 

\section{Conclusion}
In these proceedings, we presented our results for in-medium splitting rates computed using the non-perturbative collision kernel and compared them to the calculations using leading and next-to-leading order perturbative collision kernel as well as to different approximations of the in-medium splitting rates. We find that approximations to the splitting rate calculation are quite effective in reproducing the rate within their respective range of validity. 
Conversely, we observed considerable differences between the results for different collision kernels. Since the differences between the non-perturbative and the LO kernel which is usually used in phenomenological studies of jet quenching, can be on the order of $30\%$ it will be interesting to further explore the phenomenological consequences. In this context,  it is important to point out, that the collisional broadening kernel and in-medium splitting rates obtained in this study can be easily integrated into models of jet quenching using a kinetic approach or MonteCarlo simulations \cite{Yazdi:2022bru}.

\textit{Acknowledgement:}
We thank Guy D. Moore and Niels Schlusser for insightful discussions, collaboration on our previous work~\cite{Moore:2021jwe}. This work is supported in part by the Deutsche Forschungsgemeinschaft (DFG, German Research Foundation) through the CRC-TR 211 ’Strong-interaction matter under extreme conditions’– project number 315477589 – TRR 211 and the German Bundesministerium f\"{u}r Bildung und Forschung (BMBF) through Grant No. 05P21PBCAA. The authors also gratefully acknowledge computing time provided by the Paderborn Center for Parallel Computing (PC2).

\end{document}